# Adaptive Video Streaming with AI-Based Optimization for Dynamic Network Conditions

Mohammad Tarik
Computer Engineering Department
University of Mosul
Mosul,Iraq
Mohammad.t.mohammad@uomosul.edu.iq

Qutaiba ibrahim
Computer Engineering Department
University of Mosul
Mosul,Iraq
Qut1974@gmail.com

*Abstract*—The increase in video streaming has presented a challenge of handling stream request effectively, especially over networks that are variable. This paper describes a new adaptive video streaming architecture capable of changing the video quality and buffer size depending on the data and latency of streamed video. For video streaming VLC media player was used where network performance data were obtained through Python scripts with very accurate data rate and latency measurement. The collected data is analyzed using Gemini AI, containing characteristics of the machine learning algorithm that recognizes the best resolution of videos and the buffer sizes. Through the features of real-time monitoring and artificial intelligence decision making, the proposed framework improves the user experience by reducing the occurrence of buffering events while at the same time increasing the video quality. Our findings therefore confirm that the proposed solution based on artificial intelligence increases video quality and flexibility. This study advances knowledge of adaptive streaming and offers an argument about how intelligent data-driven approaches and AI may be useful tools for enhancing the delivery of video in practical environments.

## I. Introduction

As the consumption of multimedia material escalates, video streaming is taken to be a natural element of the digital environment. Services such as YouTube, Netflix, and live streaming services rely on dynamic features that fit the streaming technology to deliver continuous video playback when the network changes. Adaptive video streaming changes the stream video quality and the buffer size to match the current Network conditions to quickly provide a perfect viewer experience. However, efficiently managing video quality, data rate, latencies, and buffer has not been easily achievable yet. Fluctuating networks introduce constant quality fluctuations, buffering pauses, or decreased rates, consequently, the QoE for users[1-2].

Adaptive video streaming is one of the key components of experiences the modern multimedia consumption wherein the adaptation of high quality video content is needed within different network settings as shown in Figure 1. A not less important application of streaming media is in providing content for entertainment, education, and information; as such efficient buffering strategies to ensure that Quality of Experience (QoE) is provided has never been more important. The purpose of this paper as a whole is to analyse the features of adaptive video streaming and to highlight such aspects as a prediction of the most appropriate strategies of buffering that would further improve consumers' satisfaction as well as the reduction in interruptions during playback[3].

The use of video streaming technologies has been greatly impacted by a relatively new technique called DASH, or Dynamic Adaptive Streaming over HTTP which defines video content into segments of different bitrate. This approach helps the clients to be able to choose the best quality in response to the current network conditions to eradicate problems like buffer underruns and playback interruptions [4]. On the contrary, One of the most valuable aspects of DASH is more valuable when bandwidth is unpredictable; since it can switch between different qualities without impeding users; [5-6]. The consideration of video quality adaptation depending on the connectivity as one of the key factors eventually affects user satisfaction, especially in conditions of various connection quality, for example, in mobile networks [7-8].

The buffering operation remains an essential aspect of implementing adaptive streaming systems. Thus, more than just protection against pauses during playback, effective buffering also contributes to the increase in stream quality as perceived by end consumers. Recent studies have focused on segment request policies that are aware of bandwidth which balance QoE of the user by adapting the bitrate of video segments regarding available bandwidth [6,9]. In the same way, such strategies will help streaming services improve the playback experience, and minimize rebuffering instances that can disrupt the users' experience [10-11]. Furthermore, the use of machine learning models for predicting bandwidth variations and decision making on the best buffering strategy deemed to form a promising area regarding improving the overall adaptive streaming [12-13]. Such reconstructions can also employ data analysis to foresee network status and adapt buffer options closer to viewer expectations.

However, more analysis is needed here with the focus on the customer perspective to create effective and adaptive streaming solutions. The previous studies show that users show distinctive sensitivity profiles for buffering and quality interference that directly impact QoE [7-8]. Streaming service providers should use user behavior analysis to optimize the buffering technique for each user to increase satisfaction levels [14]. Moreover, more targeted adaptation models based on context and user parameters can be used to enhance the buffering algorithms by aiming them at the specific viewer.

The challenge of Adaptive video streaming is a complex concept and the issues connected with it have technical, behavioral as well as context based elements. Given the rising popularity of streaming services, efficient buffering techniques that can respond with proper algorithms based on network quality and user preferences will continue to be a prime area of interest. This paper will discuss the approaches and the tools used in effective buffering strategies in adaptive video streaming, in addition to reviewing the various trends advanced in the field in the present day.

As a result, the adaptation in video streaming tends to be dynamic due to clients' characteristics that make the buffering approach crucial to improving QoE. AI presents an opportunity to predict and adapt to network changes, enhancing video delivery. This paper aims to fill these gaps by presenting a framework to identify data rate and latency and adapt the video resolution and buffer size accordingly. The approach incorporates Python to analyse network statistics while streaming videos and uses Gemini Artificial Intelligence to provide recommendations based on obtained statistics. In particular, the system determines the optimal video quality depending on current link utilization and continuously manages the buffer size to reduce the impact of pauses.

This paper is organized as follows: In section 2, the artificial intelligence concerning adaptive streaming is explained. Section 3 describes the method such as setting up VLC streaming, function for data collection using Python, and integration with Gemini AI. The results are provided in Section 4 and analyse the performance of the suggested framework Depending on the networks. In the final section of the paper, section 5, the researcher provides a summative review of the studies outlined in this paper and the research directions for the future.

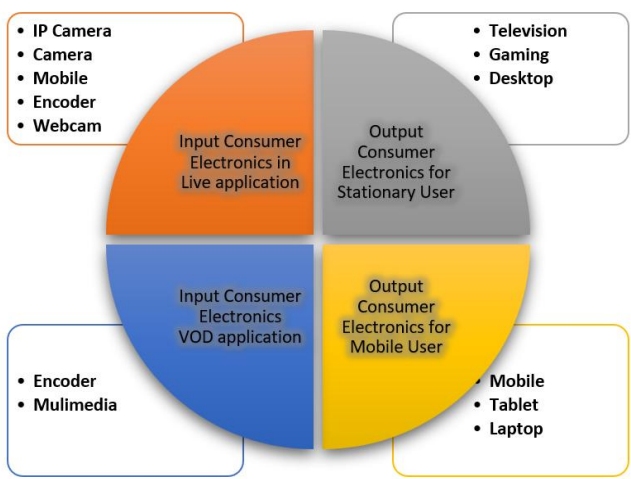

Figure 1: Video streaming applications[15].

## II. ARTIFICIAL INTELLIGENCE IN ADAPTIVE STREAMING

Adaptive video streaming is a transformative force that has been enriched by the tool of Artificial Intelligence (AI). Machine learning algorithms and data analytics can make the streaming method efficient by providing value-added advice on performance improvement, which leads to a stable, and efficient, and effective streaming platform for users [16-17]. Previous methodologies of ABR are substantially more basic, the decision making primarily driven by assessments of bandwidth with minimal consideration for other factors in the immediate environment. However, these methods could often mask important parameters like latency, or historical network performance and thus are associated with an inaccurate quality adjustment [18].

Machine learning models can therefore be used to mine huge amounts of past data and learn how the network is likely to behave [19]. With this capability, it is possible to make the right decision when it comes to changing video quality and buffer size. For example, AI algorithms hold the ability to estimate network traffic, and given a certain traffic history, the streaming service can reduce the quality of the video before it affects the consumers and brings in the buffering notice [20]. Thus, using the essence of AI in precise anticipation of the user's needs and network conditions leads to a more consistent view of the fare [21].

With AI in adaptive streaming, changes in the network conditions dictating speed, quality, and time can be in real-time[22-23]. For example, models can always track data rate and latency while changing the quality of the video depending on data results without fixed threshold values. This reliability is particularly important in the case of a live streaming event or when users switch from one network connection to another, such as a transition from Wi-Fi to cellular. Whenever decision making occurs, AI is capable of analyzing [25] numerous criteria at once excluding video quality, users' preferences, and network, in the real time [16,24].

Another important region where AI is useful is the efficient management of buffers. In using real-time data, AI is capable of discovering the right buffer size depending on the different connection conditions of a network. For example: During latencies, AI can recommend increasing the buffer size so that noise will happen but at the same time, suggest that the buffer size is reduced during periods when no such latencies are present to reduce resource usage. This dynamic adjustment also helps minimize occurrences of buffer underruns and improves cumulative user experience [17].

AI's predictive capabilities extend to identifying potential issues before they escalate into problems. By monitoring user behavior and system performance, AI can flag anomalies that may indicate network degradation. This proactive approach allows streaming platforms to implement corrective measures, ensuring a consistent and high-quality streaming experience.

## III. METHODOLOGY

This section describes the strategy applied to build the adaptive streaming framework. The methodology encompasses video streaming setup with VLC, data rate and latency acquisition with Python, and subsequently application of the AI embedded in Gemini to evaluate the precise resolution of video and the buffer size in real time.

### A. Streaming Setup with VLC

The VLC media player is used in creating an environment for video streaming. The fact that VLC supports RTSP, HTTP, and RTP/UDP streaming protocols makes it an ideal tool for developing a flexible video streaming environment. It mirrors the operating n/w in a way as it fluctuates the available bandwidth, it also mimics a crowded network thereby changing the video bit rates.

Each video stream required Steps for Streaming Setup: first, select the video content IP for streaming. Second, configure the streaming protocol, VLC can stream the video using real time streaming protocol RTSP. which provides low-latency transmission that is well suitable for use in real time video applications. Third, select the quality of the stream (Relative to output size, e.g. 360p, 720p, 1080p) based on the video server capability and the connection bitrate. fourth, the buffer size for the stream which determines the amount of data pre-loaded.

### B. Data Rate and Latency Measurement

For improving the quality of video streaming, in particular, an analytical measurement of data rate and latency in real time is crucial. These metrics are collected by using different Python scripts, and the resulting values are stored for analysis. It measures data rate (in Kbps) as to how many Kilobits per second the system transmits the video data and Latency (in milliseconds) as to how much time it takes to deliver the packets.

The data rate is therefore obtained by analysing the network traffic spanning the VLC server and the client device. In the Python script, the authors utilize the psutil and socket libraries to measure the processing of the Data collection as a core component of the adaptive video streaming system since the appropriate receivers and senders' metrics are vital for understanding the network and video quality. The following steps outline the data rate measurement process: The script starts a loop through which the functions collect the statistics of the network utilizing the psutil. This function gives the byte count for the network interfaces as both transmit and receive data. To get the data rate, the number of bytes is then divided by a time interval.

Another important parameter that influences video streaming transmission quality is latency, as it defines the time interval between sending and receiving the data packets. To evaluate latency, the Python script initiates a timer at request transmission and stops it upon response receipt, measuring the round-trip time between server and client. The latency is measured in milliseconds (ms) by subtracting the second time from the first time.

### C. AI-Based Analysis

Gemini AI, a next-generation machine learning model by Google, would be the model that can analyse different insights and give sophisticated insights. It was chosen because of the higher accuracy and integration capabilities over the traditional techniques. However, during adaptive streaming, Gemini AI analyses data rate and latency ahead of time within the stream and produces the greatest video resolution that is sustained using optimization of buffer sizes. It makes the prediction much better than preceding ABR algorithms and at some point, subsequently reduces ineffective quality swaps and buffering cases. In other words, Gemini AI operates pattern matching with machine learning where it tells what of the video stream settings should be selected and how big a buffer should be set should the current conditions of the network be such.

### D. Framework Algorithm

The concrete structure of the specific adaptive video streaming framework is based on several threads for network observing, artificial intelligence decision making, and streaming to work in an optimal way. It operates with two parallel threads: video surveillance for the first camera, and the second one dedicated to AI tracking and AI-made decisions and streaming. This is why such a separation is necessary so that high-quality monitoring and analysis complicate ordinary video playback, as shown in Figure 2.

The first of the two threads, for monitoring and the decision-making process, collect actual distinctive attributes like data rate and latency to compile those with the learned model (Gemini AI) for choosing the video definition and preferable buffer size. This is arrived at by division of the data rate with the use of the psutil library in the Python programming language and then conversion of the result to kilobits per second. At the same time, the latency level is monitored by the assessment of the Round-Trip Time (RTT) level. After the detailed analysis of the gathered data rate, latency, and time stamp, all the collected data is stored in a real-time CSV file and after analysis of the results. All this is then used to train the AI model for predicting the behaviours of the network as well as give out the ideal changes to be made to the parameters. If the data rate lowers, the quality of the video is reduced to 720 or 480 resolution or below in the case of the AI. Whenever latency goes beyond 100 ms, the buffer size must increase to avoid stalling. Under other conditions of the network, the resolution is increased to 1080p or 4K, or to the highest possible, and the buffer size is lessened in order to yield the best results. This thread continuously updates the streaming parameters and communicates them to the video streaming thread ensuring synchronization between decision-making and playback.

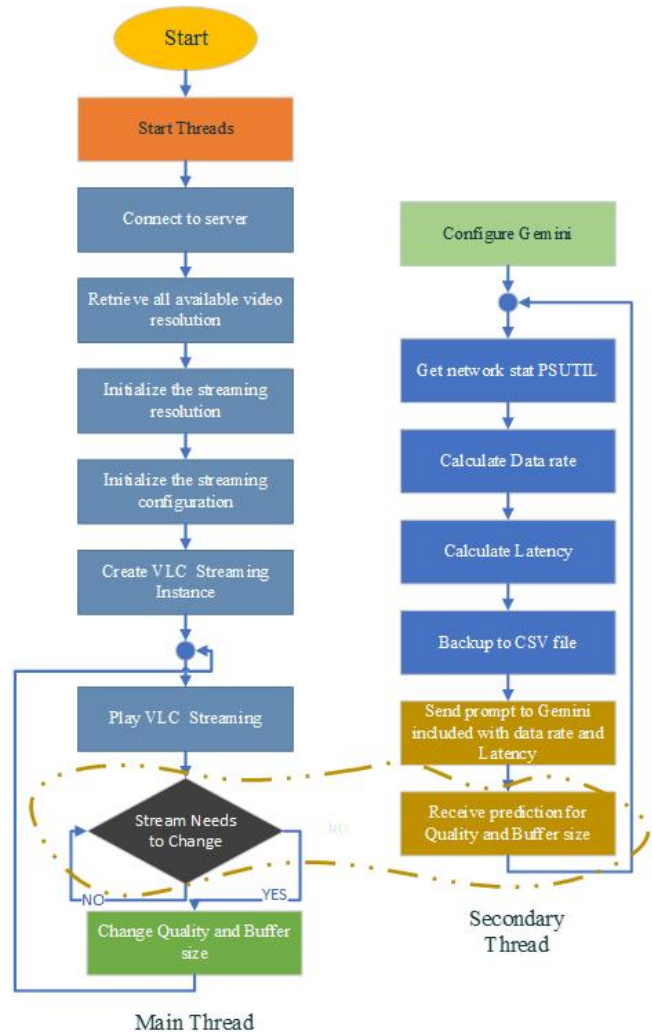

Figure 2: Proposed Framework Algorithm.

The second is devoted to starting and sustaining the video stream using VLC with the aid of a separate thread. It works independently to prevent interruption due to network checking and thus plays smoothly. First, it begins in a default quality which is the stream resolution. In this case, within the session, enables adjustment of the streaming parameters including the resolution and buffer size based on the AI prediction results in the monitoring thread. In case network conditions get worse, the streaming thread has to adapt to reduce video quality and increase the size of the buffer proportional to Gemini and vice versa. the user can change the quality directly from the streaming thread.

Regarding the data flow as shown in Figure 3, the flow is enshrined in the flowchart as follows; The main flow presents the general relationship between the user, the main thread, the secondary thread, and the relevant APIs including VLC, YouTube, and Gemini. The user starts by entering the most popular web resource – YouTube link which goes through the main thread to get available video resolutions using YouTube API (yt-dlp). The user chooses a resolution, and then the main thread sets up the stream using the VLC API at which the stream is initialized. At the same time, the main thread starts the second thread for analyzing the properties of the network such as delay, and data rate.

the results written in an Excel file. the secondary thread interacts with the Gemini API sending the average latency and data rate to get the probable optimal video resolution and the buffer size. All such predictions are recorded and then transferred to the principal thread where results are shown to the user. The user can choose the resolution on the fly easily while switching is facilitated by the main thread through the use of the VLC API. It also illustrates some of the controls the user has once within the stream, such as stopping the video's streaming and ending the monitoring process.

The multi-threaded approach enhances system effectiveness by making sure that the monitoring load does not affect the play-back of video material. Another advantage of this structure is that it responds dynamically to changes in the network conditions thereby maintaining a good view. Thanks to utilizing one of the threads for AI analysis and network traffic monitoring, the framework ensures consistent streaming in the course of complex decision making.

## IV. RESULT AND DISCUSSION

The effectiveness of the proposed adaptive video streaming framework was measured according to gains in video quality and the reduction of buffering situations while receiving video under different network conditions. Results obtained from multiple VLC streaming sessions showed two main strong relationships between network performance measurements including data rate and latency and the streaming. The main performance parameters investigated were data rate, and latency measured and fed to the AI. When implemented in environments with stable high bandwidth above 1 Mbps, the system delivered high video quality (1080p) with very limited instances of buffering episodes. On the other hand, tests done in a real environment of bandwidth conditions proved the operation of the AI-driven approach to the real environment. Figure 4 shows the ability to track the changes in the network.

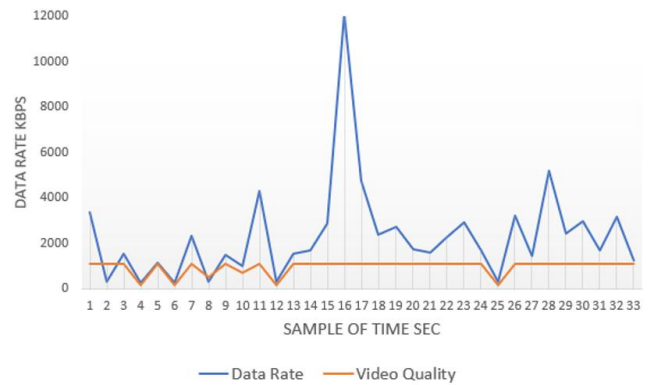

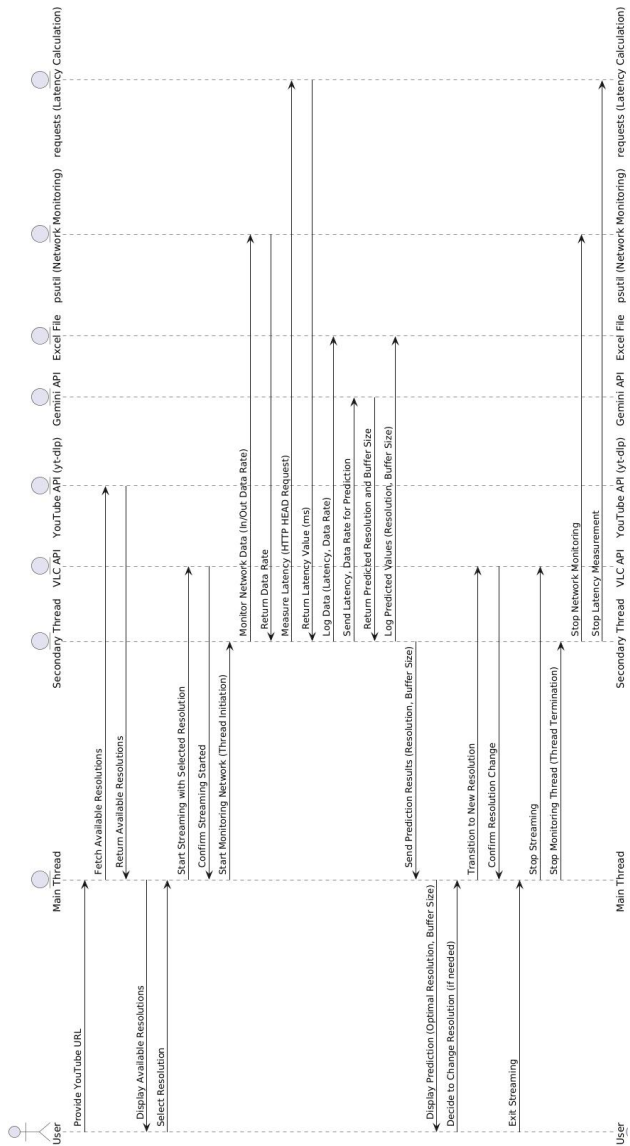

figure 3: Sequence flow diagram for the system.

The secondary thread uses the psutil library to evaluate data rates and the requests library to determine latency, with figure 4: Data rate fluctuations and corresponding AI-based resolution adjustments during network testing.

In figure 3 shows a great fluctuation in the result which is not practical to change the stream configuration every measurement period. Also, it will have a negative effect on the users and also reduce the quality of experience. To overcome this issue two approaches are considered: first, the framework will assume it is changed for a short period so it relies on the buffer to compensate for this change. Second: the framework will track the network state for three seconds and then consider the average performance and decide, is change required in the stream based on the AI prediction.

Figure 5 shows the behaviors of the framework while performing tracking for the network state.

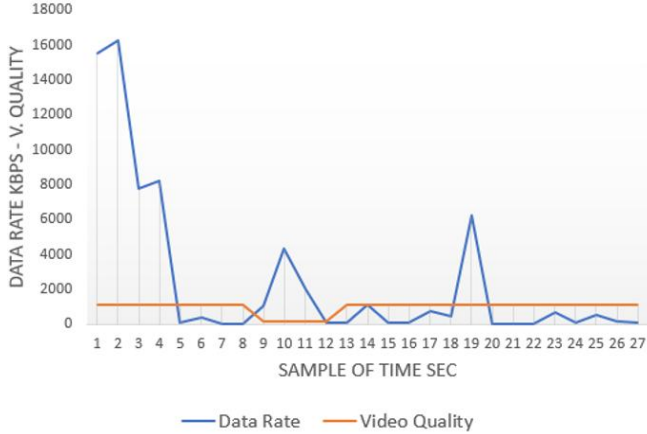

figure 5: framework algorithm's ability to deal with the network fluctuation.

From the data that was collected the data rate at peak capacity was discovered to be at its highest average of 16 Mbps but the least was only at about 0.03 Mbps of the constrained capacity. Average latency measurements fluctuated significantly, with lower average latencies achieved near 8 ms when using a stable network and a significantly higher average latency of up to 145 ms in the loaded and congestion networks. These metrics were properly interpreted by the AI part of the system. Machine learning-based algorithms used in Gemini AI help to predict online video quality and buffer size by using current network performance measures and previous data rate and latency. Figure 6 shows the prediction sample.

```
Latency: 1289.52 ms Bandwidth (In/Out): 341.71 Kbps / 27.80 Kbps
Latency: 1462.85 ms Bandwidth (In/Out): 1247.46 Kbps / 36.85 Kbps
Latency: 1526.92 ms Bandwidth (In/Out): 1880.94 Kbps / 19.88 Kbps
Latency: 1109.15 ms Bandwidth (In/Out): 1789.82 Kbps / 44.41 Kbps
the AI prdiction resolution 1080p

the AI prdiction buffer size   1355904
```

Figure 6: Framework result.

An essential outcome of the framework as shown in Figure 7 was the improvement in buffer management. The AI's ability to adjust buffer sizes dynamically based on real-time latency data contributed to a significant reduction in buffering events. For instance, when latency spikes were detected, the framework proactively increased the buffer size, preventing playback stalling. This approach facilitated a smoother viewing experience, particularly during unpredictable network conditions.

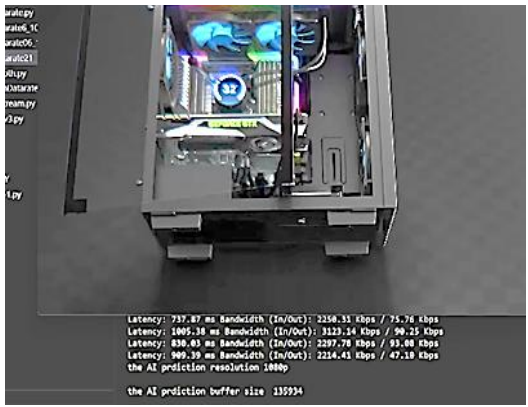

Figure 7: Framework screenshot.

Table 1 introduces a comparison of the adaptive video streaming techniques in the periods between 2022 to 2024. The initial phase of the work involved the development of Bitrate Ladders and, superimposing, Coded Unit size predictions with the help of Convolution Neural Networks, Artificial Neural Networks, and XGBoost. Recent works shifted to extending the approach for online adjustment and energy consumption minimization, where novel techniques of Random Forest and XGBoost were observed. Our approach is to combine Artificial Neural Networks with Gemini Artificial Intelligence to enable real-time and adaptive adjustment of the video resolution and the buffer size to ultimately improve the quality of experience in terms of variability in the overall network conditions.

Table 1: Comparison of methods.

| Work | Scope | Input | Artificial Intelligence | Output |
|---|---|---|---|---|
| [26] | Bitrate ladder construction | Video size, bit rate, frame rate, resolution | Convolutional Neural Network, Artificial Neural Network, XGBoost | Bitrate ladder |
| [27] | Bitrate ladder construction | Video frame, resolution, coding information | Convolutional Neural Network, Artificial Neural Network, XGBoost | Bit rate |
| [28] | Coding Unit size prediction | Motion information, residual variability information | Support Vector Machine | Coding Unit size |
| [29] | Video preset selection | Temporal information, brightness information | XGBoost | Video preset |
| [30] | Resolution and Quantization Parameter Selection | Spatial information, temporal information, frame rate, resolution, Quantization Parameter, codec type | XGBoost | Energy and Video Multimethod Assessment Fusion |
| [31] | Resolution selection | Spatial information, temporal information, bit rate | Random Forest | Encoding time, Exponential Peak Signal-to-Noise Ratio |
| Proposed Work | Resolution selection and buffer size (Adaptive video streaming) | Latency, data rate, buffer size | Gemini Artificial Intelligence | Optimal video resolution and buffer size |

V. CONCLUSION

There are strong benefits to be gained by the use of AI algorithms in conjunction with adaptive streaming technologies. AI makes streaming more efficient and user-friendly by enabling accurate decision-making, real-time adjustment, allocation of better buffer control, and optimal maintenance schedules. The findings highlighted that the integrity of the specifications should be preserved as outcomes emphasized the benefits of using AI in adaptive video streaming, with specific reference to volatile network situations. Through this framework, the work was able to show how real-time data analysis can impact user experiences stressing on reduction of interruption rates and improvement of video delivery. Due to real-time latency data used in this approach to determine buffer size, this approach can perform well in mobile and low bandwidth situations. To this end, this study shows the appropriateness of AI in dynamic video streaming scenarios to enhance the existing techniques by presenting a real-time environment that considers unpredictable network fluctuations. Further work will be aimed at increasing the efficiency of the AI algorithms to get even more precise results and extend the proposed approach to a broader range of network conditions. In subsequent work, we will also incorporate reinforcement learning algorithms to get better estimates and explore additional performance under extremely low bandwidth environments.